\documentclass[aps,twocolumn,groupedaddress,showpacs]{revtex4}
\usepackage{graphicx}
\begin{document}
\bibliographystyle{apsrev}


\title{Variation of Specific Heat with $\it{x}$ and $\it{y}$ in Na$_x$CoO$_2\cdot$$\it{y}$H$_2$O/D$_2$O}


\author{R. Jin$^1$}
\email[]{email address: jinr@ornl.gov}
\author{B.C. Sales$^1$}
\author{S. Li$^2$}
\author{D. Mandrus$^{1,2}$}

\affiliation{$^1$Condensed Matter Sciences Division, Oak Ridge
National Laboratory, Oak Ridge, TN 37831}
\affiliation{$^2$Department of Physics and Astronomy, The
University of Tennessee, Knoxville, TN 37996}


\date{\today}

\begin{abstract}
We report specific heat measurements down to 0.4 K on the layered
oxide Na$_x$CoO$_2\cdot\it{y}$H$_2$O/D$_2$O with 0 $\le$ $\it{x}$
$\le$ 0.74 and $\it{y}$ = 0 and 1.4. For the nonhydrated system
($\it{y}$ = 0), the electronic specific heat coefficient
$\gamma_N$ and the Debye temperature $\Theta_D$ vary
nonmonotonically with $\it{x}$, both displaying minima when
$\it{x}$ is close to 0.5. This indicates a systematic change of
the electronic and vibrational structures with Na content.   For
both hydrated and deuterated systems ($\it{x}$ = 0.35 and $\it{y}$
= 1.4), the specific heat reveals a sharp peak with $\Delta$C$_p
\sim$ 45.5 mJ/mol-K at T$_c^{mid} \sim$ 4.7 K and an anomaly at
T$_x \sim$ 0.8 K. While the origin of the later is unknown, the
former corresponds to the superconducting transition.
Interestingly, the electronic specific heat, after subtracting
lattice and Schottky contributions, exhibits roughly T$^2$
behavior between 0.2T$_c$ and T$_c$. This can be explained by
assuming an unconventional superconducting symmetry with line
nodes.  The results obtained under applied magnetic field further
support this scenario.
\end{abstract}
\pacs{74.20.Rp, 74.25.Bt, 74.25.Jb, 74.90.+n}
\maketitle

There is growing evidence that the strong electron-electron
correlation in layered Na$_x$CoO$_2$ is responsible for some of
its anomalous physical properties such as its ``colossal''
thermopower \cite{1,2} and possibly superconductivity in its
hydrated form.\cite{3} Although it is a good metal with high
electrical conductivity for a wide Na doping range except for
$\it{x}$ = 0.5, both local density approximation (LDA)
calculations \cite{4} and experimental work \cite{5} indicate that
the itinerant bands of Na$_x$CoO$_2$ are very narrow with $\it{W}
\ll \it{U}$, where $\it{W}$ is the band width and $\it{U}$ is the
effective on-site Coulomb interaction. This implies a high value
of the density of states (DOS) at the Fermi level. However, all
specific heat data reported so far reveal a weak or moderate
enhancement of electronic specific heat coefficient $\gamma_N$ for
both hydrated and nonhydrated cases compared to the value from LDA
band structure.\cite{4,6,7,8,9} In these reports, the $\gamma_N$
value was obtained by analyzing specific heat data above 2 K for
Na content $\it{x}$ in the range of 0.3 - 0.8. It is possible that
the extracted $\gamma_N$ value does not represent that for T
$\approx$ 0 K. On the other hand, recent calculations, using the
LDA+U method, suggest that the strength and effect of Coulomb
interactions are reduced with decreasing $\it{x}$.\cite{10} One
would thus expect the variation of $\gamma_N$ with $\it{x}$.

On the experimental side, the electronic properties of hydrated
and nonhydrated Na$_x$CoO$_2$ have not been systematically studied
as a function of composition. While the phase diagram shown in
Ref. \cite{11} is constructed from electrical transport and
magnetic measurements for 1/3 $\le \it{x} \le$ 3/4 and $\it{y}$ =
1.4, little is known about how the thermodynamic properties vary
with both $\it{x}$ and $\it{y}$. Of particular importance is the
specific heat behavior in the superconducting state of the system,
which can provide key information about the superconducting
pairing symmetry. In this Letter, we report the low-temperature
specific heat of Na$_x$CoO$_2\cdot\it{y}$H$_2$O/D$_2$O with 0 $\le
\it{x} \le$ 0.74 and $\it{y}$ = 0 and 1.4.

Single crystals of Na$_x$CoO$_2$ were used for specific heat
measurements.  Starting with Na$_{0.74}$CoO$_2$ single crystals
grown using a flux method,\cite{12} crystals with lower Na content
were obtained by chemical deintercalation as described in Ref.
\cite{6}. By controlling the deintercalation time, we obtain
single crystals with x $\approx$ 0.72 and 0.48, as determined from
measurements of the $\it{c}$-axis lattice parameter using the
calibration curve in Ref. \cite{11}.  However, it is known that
the above technique cannot extract all Na from Na$_x$CoO$_2$.  In
order to obtain CoO$_2$ ($\it{x}$ = 0) with the hexagonal
structure, we extract all Li from LiCoO$_2$ powders using
NO$_2$BF$_4$ as described in Ref. \cite{13}. Polycrystalline
samples of Na$_{0.35}$CoO$_2$ and superconducting
Na$_{0.35}$CoO$_2\cdot$1.4H$_2$O/D$_2$O were prepared following a
procedure similar to that described in Ref. \cite{14}. Specific
heat measurements were carried out using a PPMS (Physical Property
Measurement System) from Quantum Design.

In Fig.\ 1a, we present the temperature dependence of the specific
heat C$_p$ for single crystalline Na$_{0.74}$CoO$_2$ (filled
circles), Na$_{0.72}$CoO$_2$ (unfilled circles),
Na$_{0.48}$CoO$_2$ (crosses), polycrystalline Na$_{0.35}$CoO$_2$
(filled diamonds), and CoO$_2$ (unfilled diamonds). Note that
C$_p$ varies monotonically with T between 0.4 and 10 K and the
curve tends to move upward with increasing $\it{x}$. For easy
analysis, we replot the data as C$_p$/T versus T$^2$ as shown in
Fig.\ 1b. If the specific heat is due to electrons and phonons
only, it is expected that C$_p$/T will be proportional to T$^2$ at
low temperatures. While this is true for CoO$_2$ (x $\sim$ 0)
between 0.4 and 10 K, C$_p$/T clearly deviates from linearity
below $\sim$ 2 K for samples with $\it{x} \neq$ 0. The
low-temperature upturn indicates that an additional contribution
sets in which increases with decreasing T. Similar behavior was
reported previously and was attributed to a Schottky
effect.\cite{7} Thus, for T $\ll \Theta_D$ (Debye temperature), we
may describe the specific heat using
\begin{equation}
C_p/T=\gamma_N+(\beta_3 T^2+\beta_5 T^4)+\alpha
\frac{e^\frac{\Delta}{T}}{T^3(1+e^\frac{\Delta}{T})^2}.
\end{equation}
Note these terms are corresponding to contributions from
electrons, phonons (terms inside the bracket), and Schottky
anomalies,\cite{15} respectively. Here, $\beta_3$ =
N(12/5)$\pi^4$R$\Theta_D^{-3}$ with R = 8.314 J/mol-K and N =
atomic number per formula unit, $\beta_5$ is a constant describing
the anharmonic coupling, $\alpha$ is a constant proportional to
number of two-level systems, and $\Delta$ is the energy separation
between the two levels.

\begin{figure}
\includegraphics[keepaspectratio=true, totalheight =3.4 in, width =
3.3 in]{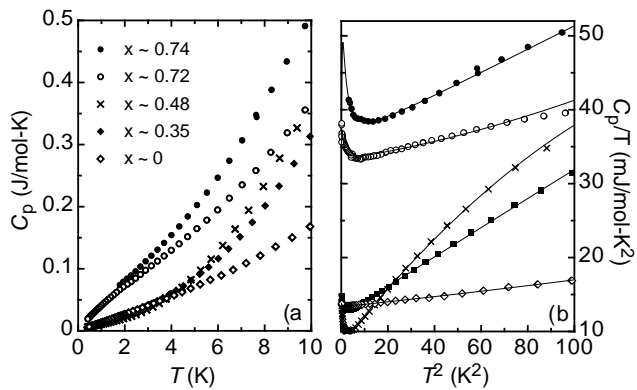} \caption{(a) Temperature dependence of the
specific heat C$_p$ for single crystal Na$_{0.74}$CoO$_2$ (filled
circles), Na$_{0.72}$CoO$_2$ (unfilled circles),
Na$_{0.48}$CoO$_2$ (crosses), polycrystalline Na$_{0.35}$CoO$_2$
(filled diamonds), and CoO$_2$ (unfilled diamonds) between 0.4 and
10 K. Shown in (b) is the replot of the data as C$_p$/T versus
T$^2$, and solid curves are the fits of experimental data to Eq.\
1. }
\end{figure}

The solid curves shown in Fig.\ 1b are fits of experimental data
between 0.4 and 10 K to Eq.\ 1. Note that Eq.\ 1 describes our
experimental data very well in the selected fitting range for all
studied compounds. Parameters obtained from the above fitting
procedure are listed in Table I. Using the relationship described
above, the Debye temperatures may be calculated from the $\beta_3$
values, and are given in Table I. In view of Table I, there are a
couple of remarkable features: (1) both $\gamma_N$ and $\beta_3$
($\Theta_D$) vary nonmonotonically with $\it{x}$, displaying
extrema for $\it{x} \sim$ 0.48; (2) $\alpha$ decreases with
decreasing $\it{x}$; (3) $\Delta$ does not vary much with
$\it{x}$. Compared to the calculated band value of $\gamma_N$ = 10
mJ/mol-K$^2$ for $\it{x} >$ 0.5 \cite{4,10} and 13 mJ/mol-K$^2$
for $\it{x} <$ 0.5\cite{10}, our $\gamma_N$ values have a factor
of 3 enhancement for $\it{x} >$ 0.48 but are very close for
$\it{x} <$ 0.48. This implies that the correlation effects are
moderate for $\it{x} >$ 0.48 and become negligible for $\it{x} <$
0.48, in very good agreement with LDA+U calculations.\cite{10} The
small $\gamma_N$ value for $\it{x} \sim$ 0.48 is likely related to
the charge ordering reported in Na$_{0.5}$CoO$_2$ at temperatures
below 52 K.\cite{11} For Na$_{0.48}$CoO$_2$, there may exist
partial charge ordering, leading to a small $\gamma_N$ value.  It
should also be mentioned that the $\gamma_N$ value for the end
compound CoO$_2$ is nonzero, consistent with the metallic behavior
reported previously.\cite{13}  Interestingly, the variation of Na
concentration in Na$_x$CoO$_2$ affects not only the DOS (reflected
by $\gamma_N$) but also $\Theta_D$ with a minimum occurring at the
same composition ($\it{x} \sim$ 0.48) as that for $\gamma_N$. This
suggests strong coupling between electrons and the lattice, and
the lattice tends to soften near $\it{x}$ = 0.5. The monotonic
decrease of $\alpha$ with $\it{x}$ reflects that the Schottky
contribution to the specific heat decreases with Na content, and
is almost negligible in the end compound CoO$_2$. This is
consistent with the trend that the low-temperature magnetic
susceptibility decreases with decreasing $\it{x}$.\cite{11} The
almost constant $\Delta$ value for different $\it{x}$ strongly
suggests that Schottky effect is mainly controlled by nuclear
moments.

\begin{table}
\includegraphics[keepaspectratio=true, totalheight = 2.5 in, width = 3.4 in]{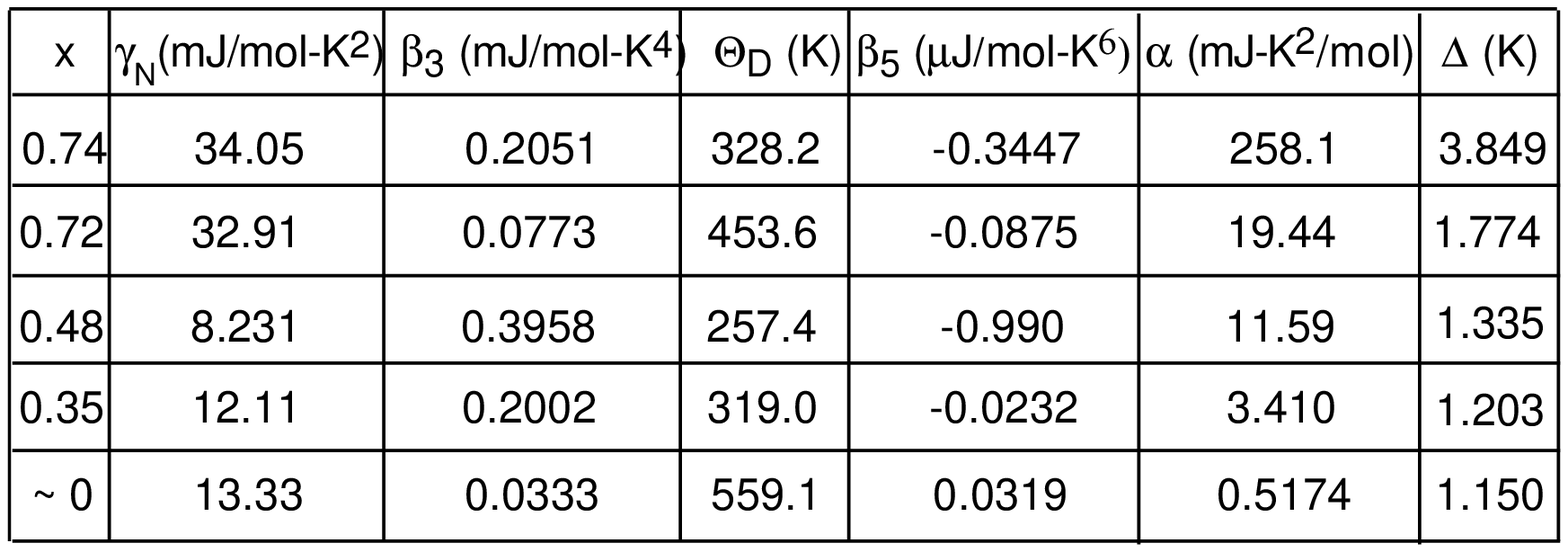}
\caption{Parameters obtained from the fit of experimental data
between 0.4 and 10 K to Eq.\ 1 and the inferred Debye temperatures
$\Theta_D$ (see text) for all x values.}
\end{table}

\begin{figure}
\includegraphics[keepaspectratio=true, totalheight =4.4 in, width =
3.3 in]{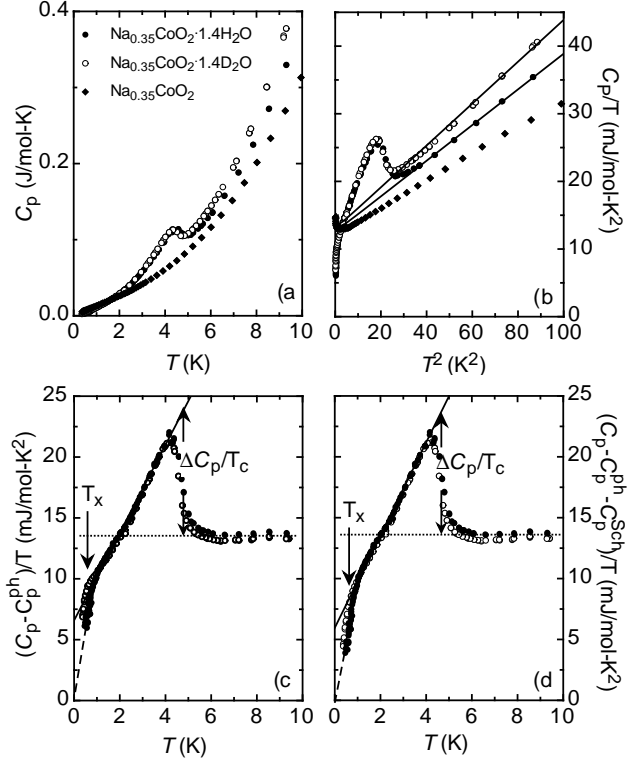} \caption{(a) Temperature dependence of the
specific heat C$_p$ for polycrystalline
Na$_{0.35}$CoO$_2\cdot$1.4H$_2$O (filled circles),
Na$_{0.35}$CoO$_2\cdot$1.4D$_2$O (unfilled circles), and
Na$_{0.35}$CoO$_2$ (filled diamonds) between 0.4 and 10 K; (b)
replot of the data in (a) as C$_p$/T versus T$^2$; solid curves
are fits of experimental data to Eq.\ 1 (see text); (c) and (d)
present the temperature dependence of (C$_p$-C$_p^{ph}$)/T (c) and
(C$_p$-C$_p^{ph}$-C$_p^{Sch}$)/T (d) of
Na$_{0.35}$CoO$_2\cdot$1.4H$_2$O (filled circles) and
Na$_{0.35}$CoO$_2\cdot$1.4D$_2$O (unfilled circles). In each case,
the solid line is the linear fit of experimental data between 0.8
and 4.1 K and the broken line is the extrapolation of experimental
data below 0.7 K to the origin.}
\end{figure}

We now focus on the specific heat of hydrated/deuturated
Na$_{0.35}$CoO$_2$.  Shown in Fig.\ 2a is the temperature
dependence of specific heat for Na$_{0.35}$CoO$_2\cdot$1.4H$_2$O
(filled circles) and Na$_{0.35}$CoO$_2\cdot$1.4D$_2$O (unfilled
circles) between 0.4 and 10 K.  Compared to that of the parent
compound (filled diamonds), the specific heat of
hydrated/deturated Na$_{0.35}$CoO$_2$ is larger at high
temperatures but smaller below $\sim$ 2 K.  Most prominent is the
specific heat anomaly below 5 K in both hydrated and deuturated
cases, reflecting the superconducting phase transition at
T$_c^{onset} \sim$ 5 K.  To obtain superconducting-state
properties, all non-electronic contributions, which are not
affected by the superconducting phase transition, should be
subtracted from the total specific heat.  As shown in Fig.\ 2b,
C$_p$/T reveals more or less linear dependence with T$^2$ above
T$_c$ for both hydrated and deuturated compounds. This suggests
that the first two terms of Eq.\ 1 are dominating. While the
low-temperature upturn is absent, the contribution from the
Schottky form may be hidden due to sharp decrease of electronic
specific heat in superconducting state. For comparison, we extract
the electronic specific heat with/without considering the Schottky
effect. Shown in Figs.\ 2c-2d are the temperature dependence of
the electronic specific heat assuming C$_p^{el}$ =
C$_p$-C$_p^{ph}$ (Fig.\ 2c) or C$_p^{el}$ =
C$_p$-C$_p^{ph}$-C$_p^{Sch}$ (Fig.\ 2d).  Here, parameters for
C$_p^{Sch}$ are assumed to be the same as that for
Na$_{0.35}$CoO$_2$ (see Table 1) since they cannot be determined
by fitting data above T$_c$, and C$_p^{ph}$(T) are obtained by
fitting experimental data between 6 and 10 K without (Fig.\ 2c) or
with (Fig.\ 2d) the consideration of the Schottky effect. It turns
out that the results are almost the same at high temperatures with
only a slight difference at low temperatures.  In both cases, note
that the substitution of D for H has little effect on the specific
heat as we concluded earlier in Ref.\cite{6}. As marked by the
dashed line in Figs.\ 2c-2d, the normal-state electronic specific
heat coefficient $\gamma_N$=13.7 mJ/mol-K$^2$ for
Na$_{0.35}$CoO$_2\cdot$1.4H$_2$O and 13.3 mJ/mol-K$^2$ for
Na$_{0.35}$CoO$_2\cdot$1.4D$_2$O, is very close to that for the
parent compound Na$_{0.35}$CoO$_2$ (see Table 1). This is in very
good agreement with a theoretical prediction that the
intercalation of water into Na$_{0.35}$CoO$_2$ results in little
change in the electronic structure.\cite{16}

\begin{figure}
\includegraphics[keepaspectratio=true, totalheight =4.4 in, width =
3.3 in]{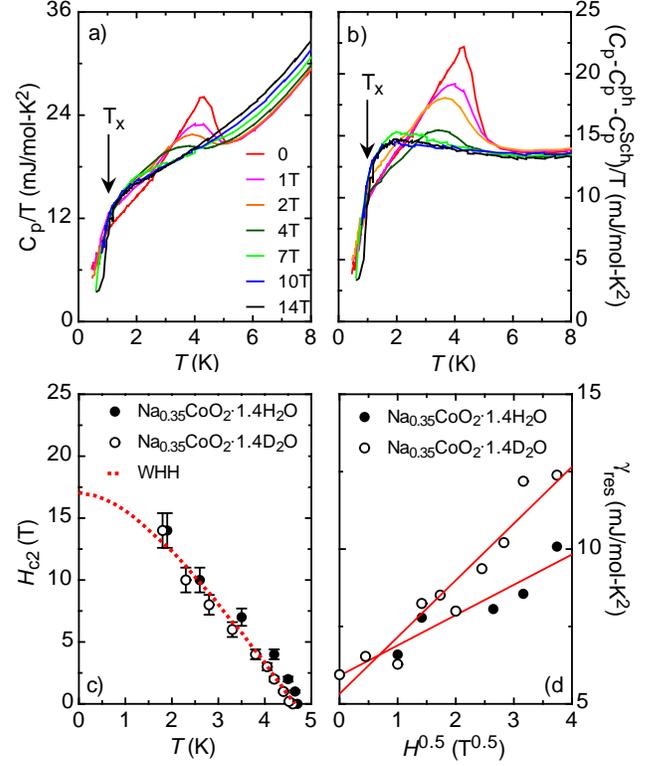} \caption{(a) Temperature dependence of the
specific heat C$_p$ for polycrystalline
Na$_{0.35}$CoO$_2\cdot$1.4H$_2$O under indicated magnetic fields;
(b) temperature dependence of the electronic specific heat with
C$_p^{el}$/T=(C$_p$-C$_p^{ph}$-C$_p^{Sch}$)/T; (c) temperature
dependence of the upper critical field H$_{c2}$ for
Na$_{0.35}$CoO$_2\cdot$1.4H$_2$O (filled circles) and
Na$_{0.35}$CoO$_2\cdot$1.4D$_2$O (unfilled circles). The dashed
line is the fit of the data to the WHH formula (see text); (d)
magnetic field dependence of the residual electronic specific heat
coefficient $\gamma_{res}$ plotted as $\gamma_{res}$ vs.
H$^{0.5}$.  The solid lines are the linear fit of the data (see
text).}
\end{figure}

Although the specific heat starts to depart from the
high-temperature behavior at 5 K and peaks at $\sim$ 4.1 K, the
thermodynamic transition temperature of
Na$_{0.35}$CoO$_2\cdot$1.4H$_2$O/D$_2$O is at T$_c^{mid} \sim$ 4.7
K where specific heat jump reaches half of the maximum. By
extrapolating the low-temperature specific heat to T$_c^{mid}$
(see Figs.\ 2c-2d), we obtain the specific heat jump $\Delta$C$_p
\sim$ 45.5 mJ/mol-K. It follows that
$\Delta$C$_p$/$\gamma_N$T$_c^{mid} \sim$ 0.71 mJ/mol-K$^2$,
roughly 50\% of the BCS value.  According to previous work, this
result is compatible with (a) 50\% superconducting volume fraction
\cite{17,18} or (b) 100\% superconductivity if there exists line
nodes in the superconducting pairing symmetry.\cite{17,19} As the
specific heat measurements are always carried out in high vacuum,
it is possible that some H$_2$O/D$_2$O is pumped out of the
material thus reducing the superconducting volume fraction. For
this reason, samples were cooled rapidly to 20 K in Helium gas
before pumping. Thermogravimetric analysis indicates that the
H$_2$O/D$_2$O content $\it{y}$ remains 1.4 after the specific heat
measurements. This and the fact that all other specific heat
measurements on either powder \cite{17,18,19} or single crystals
\cite{3} result in $\Delta$C$_p$/$\gamma_N$T$_c^{mid}$ up to 50\%
of the BCS value indicate that the small specific heat jump is
unlikely due to the sample quality.  In the second scenario, one
would expect a T$^2$ dependence of the electronic specific heat in
the superconducting state, $\it {i.e.}$ C$_p^{el}$/T $\propto$
T.\cite{20} As may be seen in Figs.\ 2c-2d, C$_p^{el}$/T varies
more or less linearly between 0.8 and 4.1 K. The solid lines are
the linear fit to the experimental data in this region. A somewhat
steeper slope is obtained when the Schottky effect is considered
(see Fig.\ 2d). Thus, it is conceivable that there exist line
nodes in superconducting pairing symmetry of
Na$_{0.35}$CoO$_2\cdot$1.4H$_2$O/D$_2$O.  However, C$_p^{el}$/T
shows an apparent deviation from linearity with sharper decrease
below T$_x$ $\sim$ 0.8 K.  This implies that not all of the
conducting electrons are involved in the superconducting phase
transition at T$_c$, leading to a small specific heat jump and a
non-zero intercept $\gamma_{res}$ when the solid lines in Fig.\
2c-2d are extrapolated to T = 0 K. Residual unpaired electrons
should be responsible for the anomaly at T$_x$, below which
C$_p^{el}$/T seems to approach linearly to the origin as guided by
the broken lines. It should be mentioned that the entropy is well
balanced when the Schottky contribution is taken into account (see
Fig.\ 2d). This strongly suggests that the sharp decrease of
electronic specific heat below T$_x$ is unlikely due to
experimental error.\cite{21}  Although the specific heat reported
by Oeschler $\it{et}$ $\it{al}$ \cite{22} was measured above 0.8
K, it was argued in Ref. \cite{22} that the temperature dependence
of specific heat of Na$_{0.35}$CoO$_2\cdot$1.4H$_2$O should behave
similarly to that of MgB$_2$ with two-gap superconductivity, where
a sharp decrease of specific heat well below T$_c$ is observed due
to a small-gap contribution.

To gain insight into the transition at T$_c$ and anomaly at T$_x$,
we have measured the specific heat of
Na$_{0.35}$CoO$_2\cdot$1.4H$_2$O and
Na$_{0.35}$CoO$_2\cdot$1.4D$_2$O under the application of magnetic
field.  Shown in Fig.\ 3a is specific heat data on
Na$_{0.35}$CoO$_2\cdot$1.4H$_2$O plotted as C$_p$/T vs. T with H =
0 (red), 1 T (pink), 2 T (orange), 4 T (forest green), 7 T
(green), 10 T (blue), and 14 T (black). Interestingly, the anomaly
at T$_x$ seems to be insensitive to H. This suggests that the
anomaly at T$_x$ is unlikely due to the small-gap contribution in
the two-gap superconductivity scenario.\cite{22} As demonstrated
in Fig.\ 3b, the anomaly remains after subtracting the phonon and
Schottky contributions. To determine whether it is related to
two-gap superconductivity or some sort of magnetic transition such
as a spin density wave, further investigation is necessary.

Nevertheless, the specific heat jump due to sperconducting phase
transition weakens and moves toward lower temperatures with
increasing H, indicating the suppression of the superconducting
transition temperature and superconducting volume fraction.  In
Fig.\ 3c, we plot the upper critical field H$_{c2}$ versus T for
Na$_{0.35}$CoO$_2\cdot$1.4H$_2$O (filled circles) and
Na$_{0.35}$CoO$_2\cdot$1.4D$_2$O (unfilled circles).  Compared to
that obtained from the electrical resistivity\cite{3} and magnetic
susceptibility measurements,\cite{23} H$_{c2}$(T) shown in Fig.\
3c varies moderately with initial slope dH$_{c2}$/dT$\mid_{T_c}$
$\sim$ -5.3 T/K. Using this value and the
Werthamer-Helfand-Hohenberg (WHH) formula for the dirty
limit,\cite{24} we may estimate the temperature dependence of
H$_{c2}$ between 0 and T$_c$. Remarkably, our experimental
H$_{c2}(T)$ follows very well the WHH expression represented by
the dashed line in Fig.\ 3c.  At T = 0 K, the WHH formula gives
H$_{c2}$(0) = 17.1 T. This allows us to estimate the coherence
length $\xi$(0) $\sim$ 44 $\AA$ using H$_{c2}$ =
$\phi_0$/2$\pi\xi^2$ ($\phi_0$ is the flux quantum). As emphasized
previously,\cite{3,23} the coherence length is short for a
superconductor with such a low T$_c$. On the other hand, the
reduction of the superconducting volume fraction due to the
application of H is reflected by the increase of $\gamma_{res}$
vaule. From linear extrapolation of C$_p^{el}$/T(T) in the linear
regime of superconducting state, we obtain $\gamma_{res}$ for each
H and plot them as $\gamma_{res}$ versus H$^{0.5}$ in Fig.\ 3d for
Na$_{0.35}$CoO$_2\cdot$1.4H$_2$O (filled circles) and
Na$_{0.35}$CoO$_2\cdot$1.4D$_2$O (unfilled circles). As guided by
the solid lines, $\gamma_{res}$ varies more or less linearly with
H$^{0.5}$ for both systems, consistent with unconventional
superconducting pairing symmetry with line nodes.

Based on the present specific heat studies, we conclude that the
electronic and vibrational properties are sensitive to $\it{x}$
but not $\it{y}$ in Na$_x$CoO$_2\cdot\it{y}$H$_2$O/D$_2$O. For the
nonhydrated system with $\it{y}$ = 0, both the electronic specific
heat coefficient $\gamma_N$ and the Debye temperature $\Theta_D$
change systematically with $\it{x}$, showing minima at $\it{x}$
$\sim$ 0.48. For the hydrated/deturated system with $\it{x}$ =
0.35, the variation of $\it{y}$ between 0 and 1.4 results in
little effect on $\gamma_N$ value. There is evidence for ``double
transitions'' with one at T$_c \sim$ 4.7 K and another at T$_x
\sim$ 0.8 K. While the origin of the later ``transition'' is
unknown, our results demonstrate that the superconducting
transition at T$_c$ is likely unconventional with line nodes in
its pairing symmetry.


\begin{acknowledgments}
We thank G.M. Veith for advice on materials preparation, and D.J.
Singh for fruitful discussions. Oak Ridge National laboratory is
managed by UT-Battelle, LLC, for the U.S. Department of Energy
under contract DE-AC05-00OR22725.
\end{acknowledgments}


%
%

%
%


\begin{references}

\bibitem{1} W. Koshibae, K. Tsutsui, and S. Maekawa, Phys. Rev. B {\bf 62}, 6869 (2000).

\bibitem{2} Y. Wang, N.S. Rogado, R.J. Cava, and N.P. Ong, Nature {\bf
423}, 425 (2003).

\bibitem{3} F.C. Chou, J.H. Cho, P.A. Lee $\it{et}$ $\it{al}$, Phys. Rev. Lett. {\bf 92},
157004 (2004).

\bibitem{4} D. J. Singh, Phys. Rev. B {\bf 61}, 13391 (2000).

\bibitem{5} B. Fisher, K.B. Chashka, L. Patlagan $\it{et}$ $\it{al}$, J. Phys.: Condens. Matter {\bf 15}, L571 (2003).

\bibitem{6} R. Jin, B.C. Sales, P. Khalifah, and D. Mandrus, Phys. Rev. Lett. {\bf 91}, 217001 (2003).

\bibitem{7} Y. Ando, N. Miyamoto, K. Segawa, T. Kawata, and I. Terasaki, Phys. Rev. B {\bf 60}, 10580 (1999).

\bibitem{8} K. Miyoshi, E. Morikuni, K. Fujiwara, J. Takeuchi, and T. Hamasaki, Phys. Rev. B {\bf 69}, 132412 (2004).

\bibitem{9} B.C. Sales, R. Jin, K.A. Affholter, P. Khalifah, G.M. Veith, and D. Mandrus, Phys. Rev. B (in press).

\bibitem{10} K.-W. Lee, J. Kunes, and W.E. Pickett, Phys. Rev. Lett. {\bf 70}, 45104 (2004).

\bibitem{11} M.L. Foo, Y. Wang, S. Watauchi $\it{et}$ $\it{al}$, Phys. Rev. Lett. {\bf 92}, 247001 (2004).

\bibitem{12} K. Fujita, T. Mochida, K. Nakamura, Jpn. J. Appl. Phys. {\bf 40}, 4644 (2001).

\bibitem{13} S. Venkatraman, and A. Manthiram, Chem. Mater. {\bf 14}, 3907 (2002).

\bibitem{14} K. Takada, H. Sakurai, E. Takayama-Muromachi $\it{et}$ $\it{al}$, Nature {\bf
422}, 53 (2003).

\bibitem{15} E.S.R. Gopal, {\it Specific Heats at Low Temperatures} (Plenum Press, New York, 1966).

\bibitem{16} M.D. Johannes, D. J. Singh, Phys. Rev. B {\bf 70}, 14507 (2004).

\bibitem{17} B. Lorenz, J. Cmaidalka, R.L. Meng, C.W. Chu, Physica C {\bf 402}, 106 (2004).

\bibitem{18} B.G. Ueland, P. Schiffer, R.E. Schaak, M.L. Foo, V.L. Miller, R.J. Cava, Physica C {\bf 402},27 (2004).

\bibitem{19} H.D. Yang, J.-Y. Lin, C.P. Sun $\it{et}$ $\it{al}$, cond-mat/0407589 (2004).

\bibitem{20} M. Sigrist, and K. Ueda, Rev. Mod. Phys. {\bf 63}, 239 (1991).

\bibitem{21} While the heat capacity of the addenda shows no anomaly at T$_x$, the specific heat
of the sample has also been carefully examined using different
methods suggested by $\it{Quantum}$ $\it{Design}$. It confirms
that the anomaly for the sample is not an artifact. For more
information about the reliability of PPMS, see J.C. Lashley
$\it{et. al}$, Cryogenics {\bf 43}, 369 (2003).

\bibitem{22} N. Oeschler, R.A. Fisher, N.E. Phillips $\it{et}$ $\it{al}$, Chinese J. Phys. (in press).

\bibitem{23} H. Sakurai, K. Takada, S. Yoshii $\it{et}$ $\it{al}$, Phys. Rev. B {\bf 68}, 132507 (2003).

\bibitem{24} N.R. Werthamer, E. Helfand, and P.C. Hohenberg, Phys. Rev. {\bf 147}, 250 (1966).

\end{references}
\end{document}